\documentstyle[pre,aps]{revtex}

\begin{document}


\title{Equilibrium and nonequilibrium applications of 
lattice-gas models in electrochemistry}

\author{
Per Arne Rikvold$^{\rm a,b,c}$
\and Gregory Brown$^{\rm a,b}$
\and M.~A.\ Novotny$^{\rm b,d}$
\and Andrzej Wieckowski$^{\rm e}$
}
\address{
$^{\rm a}$~Center for Materials Research and Technology and
Department of Physics,\\ 
Florida State University, Tallahassee, FL 32306-3016, USA\\
$^{\rm b}$~Supercomputer Computations Research Institute,\\
Florida State University, Tallahassee, FL 32306-4052, USA\\
$^{\rm c}$~Department of Fundamental Sciences, 
College of Integrated Human Studies,\\ 
Kyoto University, Kyoto 606, Japan\\
$^{\rm d}$~Department of Electrical Engineering, 
FAMU-FSU College of Engineering,\\ 
Tallahassee, FL 32310-6046, USA\\
$^{\rm e}$~Department of Chemistry and
Frederick Seitz Materials Research Laboratory,\\ 
University of Illinois, Urbana, IL 61801, USA\\
}

\maketitle

\begin{abstract}
We discuss applications of 
statistical-mechanical lattice-gas models to study static and dynamic 
aspects of electrochemical adsorption. The strategy developed to describe 
specific systems includes microscopic model formulation, calculation of
zero-temperature phase diagrams, numerical simulation of thermodynamic and
structural quantities at nonzero temperatures, and estimation of effective,
lateral interactions. We briefly review earlier work, including studies 
by Monte Carlo simulation of the adsorption of urea on Pt(100). 
As an illustrative example, we discuss in some detail recent 
applications to the underpotential deposition of Cu with sulfate on Au(111). 
Experimental and numerical results are included for equilibrium coverages 
and voltammetric currents at slow potential sweep rates, and new results 
from dynamic Monte Carlo simulations are presented. In particular, 
we present simulated results for current transients far from equilibrium, 
following sudden changes in the electrode potential. 

{~} 
\end{abstract}


\section{Introduction}
\label{sec-I}

The recent confluence of electrochemistry 
and surface science has occasioned a two-way exchange of 
experimental and theoretical techniques between these previously 
disparate disciplines. 
Here we discuss electrochemical applications of a theoretical 
method adapted from surface science: computational, 
statistical-mechanical lattice-gas modeling. 
In Secs.~2 and~3 
we briefly review some earlier equilibrium and near-equilibrium work 
\cite{RIKV88B,COLL89,RIKV91A,BLUM90,HUCK90,BLUM91}, 
including an application to the electrosorption of urea on Pt(100) 
\cite{GAMB93B,RIKV95,RIKV96}, and we briefly summarize a strategy 
to obtain effective, lateral adsorbate-adsorbate interactions by comparing 
model results with experiments performed at or near equilibrium. 
More extensive reviews of this work can be found in
Refs.~\cite{RIKV96,RIKV91B}. 

As an illustration of the approach, in Sec.~4 
we give a ``case study'' of a particular system: 
the underpotential deposition (UPD) of Cu with sulfate on Au(111) 
\cite{RIKV96,SCHU76,ZEI87,HUCK91,BLUM94A,BLUM94B,HUCK95,JZHA95B,BLUM96}. 
First we review the results of a recent Monte Carlo (MC) study 
of equilibrium coverages and voltammetric currents 
at very low potential sweep rates \cite{JZHA95B}. 
This discussion is followed by an extension of the model to study its 
{\it dynamical\/} behavior 
following a sudden change in the electrode potential. 
In particular, 
we present new simulation results for the time-dependent coverages, 
voltammetric currents, and microscopic adsorbate structures. 
The reasonable agreement with recent current-transient measurements by 
H{\"o}lzle et al.\ \cite{HOLZ94} is encouraging.

\section{Lattice-gas models of submonolayer chemisorption}
\label{sec-LGM}

The models we discuss are defined by the two-component lattice-gas 
Hamiltonian \cite{RIKV88B,COLL89,RIKV91A,GAMB93B,RIKV95,RIKV96,JZHA95B}, 
\begin{equation} 
{\cal H} 
= - \sum_n \sum_{\langle ij \rangle}^{(n)} 
         \left[ \Phi_{\rm AA}^{(n)} c_i^{\rm A} c_j^{\rm A}
         + \Phi_{\rm AB}^{(n)} \left(c_i^{\rm A} c_j^{\rm B} 
                                        + c_i^{\rm B} c_j^{\rm A} \right)  
         + \Phi_{\rm BB}^{(n)} c_i^{\rm B} c_j^{\rm B} \right] 
    + {\cal H}_3 - \sum_i \left[ \bar{\mu}_{\rm A} c_i^{\rm A}  
    + \bar{\mu}_{\rm B} c_i^{\rm B} \right] \, .
\label{eq1}
\end{equation}
Here $c_i^{\rm X}$$\in$\{0,1\} 
is the local occupation variable for species X; 
$\sum_{\langle ij \rangle}^{(n)}$ and $\sum_i$ run over all
$n$th-neighbor bonds and over all adsorption sites, respectively; 
$\Phi_{\rm XY}^{(n)}$ denotes the
effective lateral XY pair interaction through an $n$th-neighbor bond; 
$\sum_n$ runs over the interaction ranges; 
and ${\cal H}_3$ contains multi-particle interactions.
The change in electrochemical potential when one X
particle is adsorbed is $-\bar{\mu}_{\rm X}$. 
Particular systems differ in their binding-site geometries 
and the values and ranges of the lateral interactions.
The thermodynamic density conjugate to 
$\bar{\mu}_{\rm X}$ is the surface coverage,
\begin{equation}
\Theta_{\rm X} = N^{-1}\sum_i c_i^{\rm X} 
\;,
\label{eqThet}
\end{equation}
where $N$ is the number of surface unit cells. 
The mapping between the electrochemical potentials in the lattice-gas 
Hamiltonian, $\bar{\mu}_{\rm X}$, and the 
bulk activities [X] and electrode potential $E$ is accomplished through 
the relation, 
\begin{equation}
\label{eq2}
\bar{\mu}_{\rm X} = {\mu}_{\rm X}^0 
+ RT \ln \left([{\rm X}]/[{\rm X}]^0\right) - z_{\rm X}FE \;,
\end{equation}
where $R$ is the molar gas constant, 
$T$ is the absolute temperature, $F$ is Faraday's constant, 
$z_{\rm X}$ is the effective electrovalence of X, 
and ${\mu}_{\rm X}^0$ and [X]$^0$ are reference values 
which include the local binding energies. 

The interactions in Eq.~(\ref{eq1}) are {\it effective\/}
interactions mediated through several channels. 
The mechanisms include interactions between the adsorbate and the 
substrate electron structure, 
adsorbate-induced deformations of the substrate,
interactions with the fluid electrolyte, 
and (screened) electrostatic interactions. 
Theoretical and computational methods are not yet sufficiently 
advanced to obtain these multi-source interactions from first principles. 
A method to estimate them on the basis of comparison between 
lattice-gas calculations and experiments is summarized in the next section.

\section{Voltammetry and adsorbate structure near equilibrium}
\label{sec-EQ}

In the limit of vanishing potential sweep rate, 
the voltammetric current per unit cell 
is the time derivative of the charge transported across the interface 
during the adsorption/desorption process: 
\begin{equation}
\label{eq2c}
q = -e(z_{\rm A} \Theta_{\rm A} + z_{\rm B} \Theta_{\rm B}) \;,
\end{equation}
where $e$ is the elementary charge. 
The current density 
is easily obtained in terms of the lattice-gas response functions, 
${\partial \Theta_{\rm X}}/{\partial \bar{\mu}_{\rm Y}}$, as 
\begin{equation}
i = e {F}
\left\{ z_{\rm A}^2 \frac{\partial \Theta_{\rm A}}{\partial
\bar{\mu}_{\rm A}}
+ 2 z_{\rm A} z_{\rm B}
    \frac{\partial \Theta_{\rm B}}{\partial \bar{\mu}_{\rm A}}
           + z_{\rm B}^2
    \frac{\partial \Theta_{\rm B}}{\partial \bar{\mu}_{\rm B}}
    \right\} \frac{{\rm d}E}{{\rm d}t} \;.
\label{eq2b}
\end{equation}

A strategy has been developed to determine the effective interactions 
\cite{RIKV95,RIKV96}, 
providing a practical alternative to ``first-principles'' 
methods. This consists in 
fitting thermodynamic and structural model predictions from 
zero-temperature phase diagrams and numerical 
simulations at room temperature to experiments, considering 
a variety of physical quantities. 
Obviously, this also has its  
problems. The potential number of parameters is large, and there is no 
guarantee that a minimal set of interactions is unique.
Nevertheless, previous lattice-gas studies of electrochemical systems 
indicate that when 
all available experimental information is consistently included, this 
approach has considerable predictive power. 
The steps in the modeling strategy can be summarized as follows.
\vspace*{-0.2truecm}
\begin{enumerate}
\topsep=0pt
\parskip=0pt
\itemsep=0pt
\item
Use prior theoretical and experimental knowledge about the substrate
lattice structure and lattice constant and the shapes and sizes of the
adsorbate particles to formulate a specific lattice-gas model.
\item
Use experimental information about adsorbate coverages 
and adlayer structure to determine the adsorbate phases.
\item
Perform a group-theoretical ground-state calculation to 
construct a ground-state diagram (zero-temperature phase diagram) 
and determine a minimal set of
effective interactions consistent with the observed adsorbate phases.
\item
At nonzero temperatures, the thermodynamic and structural properties of
the lattice-gas model constructed through steps 1--3 can be studied
by various analytical and numerical methods, depending
on the quantities of interest and the complexity of the Hamiltonian. 
In the applications discussed here, we have mostly used numerical 
MC simulation. 
This method has the advantage that it is quite accurate, even for 
two-dimensional systems \cite{RIKV93D}, 
and relatively straightforward to implement. 
\item
The finite-temperature properties
obtained in Step 4 should be used to refine the effective
interactions by comparison with the available experiments,
or by obtaining additional experimental data for such comparison.
\end{enumerate}
\vspace*{-0.3truecm}
Steps 4 and 5 should be iterated
until satisfactory agreement between model and experiment is achieved.

Examples of the application of the approach sketched above to the 
chemisorption of small adsorbate particles on single-crystal electrodes
include the adsorption of urea on Pt(100) from perchloric acid 
\cite{GAMB93B,RIKV95,RIKV96} 
and the UPD of Cu on Au(111) from sulfuric acid 
\cite{HUCK91,JZHA95B,BLUM96}. 
Both systems exhibit a prominent peak sharpening 
in the cyclic voltammogram (CV)
when a small concentration of the adsorbate (urea or copper ions) 
is added to the electrolyte. 
Whereas the urea/Pt(100) system develops a single CV peak \cite{RUBE91}, 
in copper UPD  two peaks are exhibited \cite{SCHU76,ZEI87}. 
The voltammetric changes are much weaker or absent when the same 
substances are adsorbed on other crystal planes of the same metals, 
indicating that they depend crucially on the geometric fit between 
the adsorbates and the surface. 
These peaks are associated with adsorbate phase transitions 
\cite{ZEI87,RIKV92}. 

Below we consider in detail simulations of 
some equilibrium and nonequilibrium aspects of the UPD of Cu with sulfate 
on Au(111).

\section{UPD of Cu with sulfate on Au(111)}
\label{sec-UPD}

\subsection{Near-equilibrium aspects}
\label{sec-UPD-eq}

In underpotential deposition (UPD), a monolayer
of one metal is electrochemically adsorbed onto another in a range of
electrode potentials more positive than those where bulk deposition occurs. 
The UPD of Cu on Au(111) in sulfuric acid 
has been intensively studied, both experimentally 
(see discussion of the literature in Ref.~\cite{JZHA95B}) 
and theoretically 
\cite{HUCK91,BLUM94A,BLUM94B,HUCK95,JZHA95B,BLUM96}. 
The most striking feature in CV experiments 
on this system 
is the appearance of two peaks, 
100$\sim$150 mV apart, upon addition of Cu$^{2+}$ ions 
to the sulfuric-acid electrolyte \cite{SCHU76,ZEI87}. 
Typical CV profiles are shown in Fig.~\ref{figCVb}(a), together with
simulation results. 
In the potential range between the peaks, the adsorbate layer 
has a $(\sqrt3$$\times$$\sqrt3)$ structure 
with 2/3$\,$ML Cu and 1/3$\,$ML sulfate 
\cite{HUCK91,BLUM94A,BLUM94B,HUCK95,JZHA95B,BLUM96,TONE95}, 
first proposed by Huckaby and Blum (HB)
\cite{HUCK91}. 
Typical experimental and simulated coverages are shown versus 
electrode potential in Fig.~\ref{figCVb}(b). 

The lattice-gas model used by Zhang et al.\ \cite{JZHA95B} has interactions 
through fourth-nearest neighbors (see also Ref.~\cite{BLUM96}). 
Sulfate is assumed to coordinate 
the triangular Au(111) surface through three of its
oxygen atoms, with the fourth S-O bond away from the surface. 
This gives a triangular ``footprint'' 
that reasonably matches the Au(111) unit cell. 
The adsorption sites for the Cu and sulfate are assumed to lie
on the same triangular lattice.
The model is illustrated in Fig.~\ref{figLATb}. 

Adsorption isotherms and CV currents at room temperature 
were obtained from MC heat-bath 
simulations on $L$$\times$$L$ triangular lattices with $L$=30 and 45. 
The simulated quantities were adjusted to obtain overall agreement with 
the experimental data as described in Sec.~\ref{sec-EQ}. 
The experimental and simulated CV currents and coverages are shown 
together in Figs.~\ref{figCVb}(a) and~\ref{figCVb}(b), respectively. 
The resulting estimates for the effective lateral interactions 
are given in the caption of Fig.~\ref{figLATb}. 

It has been experimentally observed 
\cite{ZSHI94C,ZSHI95} 
that sulfate is adsorbed on top of the Cu monolayer 
at negative potentials. Zhang et al.\ used a simple mean-field 
approximation for the sulfate coverage in this second layer: 
$\Theta_{\rm S}^{(2)} = 
\alpha\Theta_{\rm C}(\frac{1}{3}-\Theta_{\rm S})$,
where $\alpha$ is a phenomenological constant.
Since the transfer of sulfate between the gold and copper
surfaces does not
involve an oxidation/reduction process, the total
charge transport
per unit cell during the adsorption/desorption
process becomes
\begin{equation}
\label{eqQ2}
q = -e [ z_{\rm S} ( \Theta_{\rm S} + \Theta_{\rm S}^{(2)} )
+ z_{\rm C} \Theta_{\rm C} ] \;,
\end{equation}
which modifies the CV current given in Eq.~(\ref{eq2b}) in a 
straightforward way \cite{JZHA95B}. 
The effective electrovalences, $z_{\rm S}$ and $z_{\rm C}$,
must be determined from experiments. 
Zhang et al.\ obtained the values,
$z_{\rm C}$$\approx$+1.7 and $z_{\rm S}$$\approx$$-$1.1 \cite{JZHA95B},
using data from Omar {et al.} \cite{OMAR93}. 
To within the statistical errors, 
these estimates agree with independent experimental 
results by Shi and Lipkowski \cite{ZSHI94C,ZSHI95}. 
Comparing with values proposed by Blum et al.\ \cite{BLUM96} based on 
a theoretical mean-field calculation, we observe that their estimate for 
$z_{\rm C}$ is about 35\% less positive 
than those obtained by Zhang et al.\ and Shi and Lipkowski, 
while there is no statistically significant disagreement about $z_{\rm S}$. 

The ground-state diagram corresponding to the interactions obtained
in Ref.~\cite{JZHA95B} is shown in Fig.~\ref{figGSb}.
For large negative $\bar{\mu}_{\rm S}$,
only copper adsorption is possible. Similarly, in the limit of large positive
$\bar{\mu}_{\rm S}$ and large negative $\bar{\mu}_{\rm C}$, the
zero-temperature phase is the $(\sqrt{3}\!\times\!\sqrt{3})_0^{1/3}$ sulfate
phase [we denote phases as 
($X\!\times\!Y$)$_{\Theta_{\rm C}}^{\Theta_{\rm S}}$]
characteristic of the hard-hexagon model corresponding to the
infinitely repulsive nearest-neighbor sulfate-sulfate interaction
$\Phi_{\rm SS}^{(1)}$
\cite{BLUM91,HUCK91,BLUM94A,BLUM94B}.
(See Fig.~\ref{figUPDconf}.)

The potential scan path corresponding to the CV
profile and coverages shown in
Fig.~\ref{figCVb} is indicated by the dotted line labeled ``1''
in the ground-state diagram, Fig.~\ref{figGSb}. 
Starting from the negative end, we consider a scan proceeding in the
direction of positive electrode potential (upper left to lower right in
Fig.~\ref{figGSb}). Near the CV peak at approximately
200~mV vs.\ Ag/AgCl (labeled Peak \#~2 in Fig.~\ref{figCVb}), the
sulfate begins to compete with copper for the
adsorption sites, resulting
in a third of the copper desorbing and being replaced by sulfate.
Due to the strong effective 
attraction between the copper and sulfate adparticles, the mixed
$(\sqrt{3}\!\times\!\sqrt{3})_{2/3}^{1/3}$ phase is formed
(see Fig.~\ref{figUPDconf}). In this phase, which
extends through the entire potential region between the two CV peaks, 
Cu forms a honeycomb lattice with a sulfate molecule at 
the center of each cell. 
As the CV peak at approximately 300~mV 
(labeled Peak \#~1 in Fig.~\ref{figCVb}) is reached,
most of the copper is desorbed within a narrow potential range.
As it is thus deprived of the stabilizing influence
of the coadsorbed copper, the sulfate is almost completely desorbed.

Dynamical aspects of the phase transition at Peak \#~1 
are discussed in Sec.~\ref{sec-FFE} below. 

\subsection{Current transients far from equilibrium} 
\label{sec-FFE}

Whereas the preceding discussion presents a near-equilibrium theory, 
with voltammetric currents at slow potential sweep rates 
obtained from adsorption isotherms, 
analysis of fast sweeps or potential-step experiments 
requires a true nonequilibrium treatment. In the early-time regime, 
mean-field rate equations can be satisfactory \cite{AVRAMI,BOSC81}. 
However, for later times, when coalescence of adsorbate islands may become 
important, this approach is no longer 
necessarily reliable \cite{RAMO96,RIKV96B,RAMO97}. For 
systems in which the phases involved are ordered, the 
microscopic adlayer structure and the dynamical 
details of the adsorption and lateral diffusion processes 
become important, even at early times. 
We have therefore initiated a study of the dynamics of electrochemical 
adsorption by numerical simulation of microscopic models. 
Here we present some preliminary results of this work. 

We consider the model for UPD of Cu with sulfate on Au(111), discussed 
in Sec.~\ref{sec-UPD-eq} above, with the parameters that were obtained 
by Zhang et al.\ through comparison with near-equilibrium 
experiments \cite{JZHA95B}. 
We focus on the current transients following sudden changes 
of the electrode potential across the discontinuous phase transition at 
Peak \#~1, and we discuss our results in the light of recent experiments by 
H{\"o}lzle et al.\ \cite{HOLZ94}. 

The model dynamics is implemented through a refusal-free dynamic MC algorithm 
\cite{BORT75,NOVO94A}, in which the transition rates 
are determined by free energies of activation (``free-energy barriers'') 
\cite{BARD80}. The transition rate from a microscopic configuration $a$ 
to a different configuration $b$ is taken to be
\begin{equation}
k_{a \rightarrow b} = 
k'\exp{\left(-\frac{\Delta G_{a \rightarrow b}^*}{RT}\right)} \;,
\label{eq:kab}
\end{equation} 
where $k'$ is a standard rate and $\Delta G_{a \rightarrow b}^*$ 
is the free energy of activation for the particular transition. 
The eventual approach to equilibrium is ensured through 
the detailed-balance condition, 
\begin{equation}
\Delta G_{a \rightarrow b}^* - \Delta G_{b \rightarrow a}^* 
= {\cal H}(b) - {\cal H}(a) \;, 
\label{eq:detbal}
\end{equation} 
where ${\cal H}(a)$ and ${\cal H}(b)$ are the values of the lattice-gas 
Hamiltonian in Eq.~(\ref{eq1}), corresponding to the two configurations. 
In these exploratory simulations we have taken the free-energy  
barrier associated with adsorption of particles of 
species X to depend linearly 
on the electrochemical potential $\bar{\mu}_{\rm X}$ (and thus on the 
electrode potential) through a Butler-Volmer type relation 
\cite{HOLZ94,BARD80}, 
\begin{equation}
\Delta G_{\rm ads \, X}^*(\bar{\mu}_{\rm X}) = 
\Delta G_{\rm ads \, X}^*(0) - \alpha \bar{\mu}_{\rm X} \;,
\label{eq:ButVol} 
\end{equation} 
and for simplicity we set the transfer coefficient $\alpha$=1/2. 
The barriers associated with lateral diffusion are taken to be independent 
of the electrochemical potentials and, 
except insofar as they must satisfy Eq.~(\ref{eq:detbal}), 
of the local geometry of the adsorbate layer. 

The time unit used in the simulations, called MC steps per site (MCSS), 
is determined by the standard rate constants and is expected to be 
proportional to the physical time. The proportionality constant must 
be determined by physico-chemical arguments and comparison with 
the experimentally observed timescales. 

We simulated positive-going and negative-going 
potential-step experiments along the scan path labeled 
``3'' in Fig.~\ref{figGSb}, where Peak \#~1 is centered at approximately 
260~mV vs.\ Ag/AgCl \cite{JZHA95B}. 
Before the positive-going steps, the model was equilibrated in the ordered 
$(\sqrt{3} \times \sqrt{3})_{2/3}^{1/3}$ phase, 15~mV on the negative
side of the transition. At time 
$t$=0 the potential was changed instantaneously to a final value a distance 
$\Delta E$ past the transition. Following H{\"o}lzle et al., we used 
$\Delta E =$ 24, 28, 31, and 33~mV. 
{}For the negative-going steps we 
equilibrated in the disordered low-coverage phase, 
45~mV on the positive side of the transition, followed at $t$=0 
by a step into the ordered-phase region with $\Delta E = -15$~mV. 
Since the second-layer sulfate coverage is negligible in this potential 
region [see Fig.~\ref{figCVb}(b)], in this study we only considered 
first-layer effects. 

The simulated current transients [in units of elementary charges per 
Au(111) unit cell and MCSS] are shown in Fig.~\ref{figQuenchCurrent}. 
We used systems of size $60 \times 60$ with periodic boundary conditions. 
The results were averaged over 100 trials for each value of $\Delta E$, and  
the current data were smoothed using a running average over seven time points. 
As we discuss in detail below, 
the experimentally observed asymmetry with respect to 
potential steps of opposite sense is captured by our dynamical model.

The current transients for the positive-going potential 
steps, which are shown in Fig.~\ref{figQuenchCurrent}(a), 
are characterized by a sharp current spike at early times, followed later by 
a second, broader maximum.
The experimentally observed trends with respect to changes in $\Delta E$ 
are also reproduced, both for the locations and the sizes of the peaks. 
{}For the negative-going steps, no such two-peaked structure is observed. 
Instead, like in the experiments, a monotonically decreasing current is 
seen. This behavior is shown in Fig.~\ref{figQuenchCurrent}(b). 

The surface coverages, $\Theta_{\rm C}$ and $\Theta_{\rm S}$, are shown 
versus time for positive-going steps 
in Figs.~\ref{figPQuenchCoverage}(a) and~\ref{figPQuenchCoverage}(b), 
respectively. Comparison of these results with Fig.~\ref{figQuenchCurrent}(a) 
shows that the early-time current peak is almost entirely due to 
desorption of Cu. Only as $\Theta_{\rm C}$ reaches a ``metastable'' value 
in the 0.4--0.5~ML range, does the sulfate start to 
desorb at a significant rate. 
At later times the relaxation appears to follow the 
nucleation-and-growth dynamics characteristic of the decay of a metastable 
phase \cite{HOLZ94,AVRAMI,RIKV94}. 
The evolution of the spatial structure with time is illustrated by 
MC ``snapshots'' in Fig.~\ref{figSnapPlus}. The 
growth of islands of the equilibrium low-coverage phase is quite apparent. 
A detailed investigation of the 
time dependence in the late-time regime, 
including the effective ``Avrami exponent'' \cite{HOLZ94,AVRAMI} 
and its dependence on the free-energy barriers, will be 
presented at a later date \cite{BROW97B}. 

The situation following a negative-going step is quite different. 
As seen in Fig.~\ref{figNQuenchCoverage}, 
Cu is rapidly adsorbed to a coverage of approximately 1/3~ML, 
which produces the large current at early times. The snapshots in 
Fig.~\ref{figSnapMinus} [in particular Fig.~\ref{figSnapMinus}(b)] 
indicate that this structure is globally disordered, but {\it locally\/} 
has the symmetry of the Cu-only $(\sqrt{3} \times \sqrt{3})_{1/3}^0$ phase. 
{}For the particular potential steps used here, the energy of this phase 
lies between those of the mixed phase and the disordered low-coverage phase.
The subsequent approach towards the mixed, ordered equilibrium phase requires 
the removal of a large number of domain walls 
by lateral diffusion and is consequently very slow. 
Again, a detailed investigation of this dynamics and its dependence on the 
model parameters will be presented later \cite{BROW97B}.

\section{Conclusions} 
\label{sec-CON}

In this paper we have briefly reviewed some earlier applications of 
statistical-mechanical lattice-gas models to electrochemical adsorption 
problems. This approach, which was originally transferred from ``traditional'' 
surface science, has proven to be a useful and versatile tool in 
the study of multicomponent chemisorption in electrochemical systems 
near thermodynamic equilibrium. 

Following this brief review, we presented preliminary results from 
a new application of the 
lattice-gas approach to dynamical phenomena far from  
equilibrium. In particular, we considered a dynamic extension of a model 
for UPD of Cu with sulfate on Au(111), originally introduced 
by Huckaby and Blum \cite{HUCK91}. We used an extension of this model, with 
parameters obtained by Zhang et al.\ \cite{RIKV96,JZHA95B} through 
comparison of simulation results to experiments at or near equilibrium.  
To this we added activated 
dynamics including adsorption, desorption, and lateral diffusion. 
Our simulated results for current transients in this system show encouraging, 
qualitative agreement with recent experiments by H{\"o}lzle et al.\ 
\cite{HOLZ94}. This agreement includes the prominent asymmetry between the 
transients observed in positive-going and negative-going potential-step 
experiments. Although our results leave many details for 
further study, we believe they open 
the door towards numerical investigations of a wide range of interesting 
dynamical phenomena in electrochemical systems.

\section*{Acknowledgments} 
\label{sec-Ack}

Research at Florida State University was supported by 
the Center for Materials Research and Technology 
and 
by the Supercomputer Computations Research Institute
(under U.S.\ Department of Energy Contract No.\ DE-FC05-85-ER25000) 
and by U.S.\ 
National Science Foundation grants No.~DMR-9315969 and DMR-9634873. 
P.A.R.'s stay at Kyoto University during part of the work reported here 
was supported by the Japan Foundation Center for Global 
Partnership Science Fellowship Program through U.S.\ 
National Science Foundation Grant No.\ INT-9512679. 
Research at The University of Illinois was supported by U.S.\ 
National Science Foundation grant No.~CHE-9411184 and by the Frederick Seitz
Materials Research Laboratory 
(under U.S.\ Department of Energy Contract No.\ DE-AC02-76-ER01198).


\newpage

\section*{References}
\begin{thebibliography}{10}

\bibitem{RIKV88B}
P.A.\ Rikvold, J.B.\ Collins, G.D.\ Hansen and J.D.\ Gunton,
Surf.\ Sci., 203 (1988) 500.

\bibitem{COLL89}
J.B.\ Collins, P.~Sacramento, P.A. Rikvold and J.D.\ Gunton,
Surf.\ Sci., 221 (1989) 277.

\bibitem{RIKV91A}
P.A.\ Rikvold and M.R.\ Deakin,
Surf.\ Sci., 294 (1991) 180.

\bibitem{BLUM90}
L.~Blum, Adv.\ Chem.\ Phys., 78 (1990) 171.

\bibitem{HUCK90}
D.A.\ Huckaby and L.~Blum,
J.\ Chem.\ Phys., 92 (1990) 2646.

\bibitem{BLUM91}
L.~Blum and D.A.\ Huckaby,
J.\ Chem.\ Phys., 94 (1991) 6887.

\bibitem{GAMB93B}
M.~Gamboa-Aldeco, P.~Mrozek, C.K.\ Rhee, A.~Wieckowski, P.A.\ Rikvold and
Q.~Wang, Surf.\ Sci.\ Lett., 297 (1993) L135.

\bibitem{RIKV95}
P.A.\ Rikvold, M.~Gamboa-Aldeco, J.~Zhang, M.~Han, Q.~Wang, H.L.\ Richards
and A.~Wieckowski, Surf.\ Sci., 335 (1995) 389.

\bibitem{RIKV96}
P.A.\ Rikvold, J.~Zhang, Y.-E.\ Sung and A.~Wieckowski, 
Electrochim.\ Acta, 41 (1996) 2175.

\bibitem{RIKV91B}
P.A.\ Rikvold, Electrochim.\ Acta, 36 (1991) 1689.

\bibitem{SCHU76}
J.W.\ Schultze and D.~Dickertmann,
Surf.\ Sci., 54 (1976) 489.

\bibitem{ZEI87}
M.~Zei, G.~Qiao, G.~Lempfuhl and D.M.\ Kolb,
Ber.\ Bunsen Ges.\ Phys.\ Chem., 91 (1987) 3494.

\bibitem{HUCK91}
D.A.\ Huckaby and L.~Blum,
J.\ Electroanal.\ Chem., 315 (1991) 255.

\bibitem{BLUM94A}
L.~Blum and D.A.\ Huckaby,
J.\ Electroanal.\ Chem., 375 (1994) 69.

\bibitem{BLUM94B}
L.~Blum, M.~Legault and P.~Turq,
J.\ Electroanal.\ Chem., 379 (1994) 35.

\bibitem{HUCK95}
D.A.\ Huckaby and L.~Blum,
Langmuir, 11 (1995) 4583.

\bibitem{JZHA95B}
J.~Zhang, Y.-S.\ Sung, P.A.\ Rikvold and A.~Wieckowski,
J.\ Chem.\ Phys., 104 (1996) 5699.

\bibitem{BLUM96}
L.\ Blum, D.A.\ Huckaby and M.\ Legault,
Electrochim.\ Acta, 41 (1996) 2207.

\bibitem{HOLZ94}
M.H.\ H{\"o}lzle, U.\ Retter and D.~M.\ Kolb,
J.\ Electroanal.\ Chem., 371 (1994) 101.

\bibitem{RIKV93D}
A comparison of results of perturbative and nonperturbative
  calculations for one and the same two-dimensional lattice-gas model is given
  by P.A.\ Rikvold and M.A.\ Novotny, 
in L.~Blum and F.B.\ Malik (Eds.), 
Condensed Matter Theories, Vol.~8, Plenum, New York, 1993, p~627. 

\bibitem{RUBE91}
M.~Rubel, C.K.\ Rhee, A.~Wieckowski and P.A.\ Rikvold,
J.\ Electroanal.\ Chem., 315 (1991) 301.

\bibitem{RIKV92}
P.A.\ Rikvold and A.~Wieckowski, Phys.\ Scr., T44 (1992) 71. 

\bibitem{TONE95}
M.F.\ Toney, J.N.\ Howard, J.~Richer, G.L.\ Borges, J.G.\ Gordon~II, O.R.\
Melroy, D.~Yee and L.B.\ Sorenson,
Phys.\ Rev.\ Lett., 75 (1995) 4472.

\bibitem{ZSHI94C}
Z.~Shi and J.~Lipkowski,
J.\ Electroanal.\ Chem., 365 (1994) 303.

\bibitem{ZSHI95}
Z.~Shi, S.~Wu and J.~Lipkowski,
Electrochim.\ Acta, 40 (1995) 9.

\bibitem{OMAR93}
I.H.\ Omar, H.J.\ Pauling and K.~J{\"u}ttner,
J.\ Electrochem.\ Soc., 140 (1993) 2187.

\bibitem{AVRAMI}
M.\ Avrami, J.\ Chem.\ Phys., 7 (1993) 1103; 8 (1940) 212; 9 (1941) 177.

\bibitem{BOSC81}
E.\ Bosco and S.K.\ Rangarajan,
J.\ Chem.\ Soc. Faraday Trans.\ 1, 77 (1981) 1673.

\bibitem{RAMO96}
R.A.\ Ramos, P.A.\ Rikvold and M.A.\ Novotny,
in Z.~Fisk, L.~Gor'kov, D.~Meltzer and R.~Schrieffer (Eds.),
Physical Phenomena at High Magnetic Fields II, 
World Scientific, Singapore, 1996, p.~380. 

\bibitem{RIKV96B}
P.A.\ Rikvold, A.\ Wieckowski and R.A.\ Ramos,
Mater.\ Res.\ Soc.\ Symp.\ Proc.\ Ser., (1997), in press.

\bibitem{RAMO97}
R.A.\ Ramos, S.W.\ Sides, P.A.\ Rikvold and M.A.\ Novotny,
in preparation.

\bibitem{BORT75}
A.B.\ Bortz, M.H.\ Kalos and J.L.\ Lebowitz,
J.\ Comput.\ Phys., 17 (1975) 10.

\bibitem{NOVO94A}
M.A.\ Novotny,
Computers in Physics, 9 (1995) 46.

\bibitem{BARD80}
See, e.g., A.J.\ Bard and L.R.\ Faulkner,
Electrochemical Methods: Fundamentals and Applications, 
Wiley, New York, 1980.

\bibitem{RIKV94}
P.A.\ Rikvold and B.M.\ Gorman,
in D.~Stauffer (Ed.), Annual Reviews of Computational Physics I, 
World Scientific, Singapore, 1994, p.~149. 

\bibitem{BROW97B}
G.\ Brown, P.A.\ Rikvold, M.A.\ Novotny and A.\ Wieckowski,
in preparation.

\end{thebibliography}

\newpage


\section*{Figure Captions}
\begin{figure}[]
\caption[]{
UPD of Cu with sulfate on Au(111). 
Electrolyte composition: 1.0~mM CuSO$_4$ + 0.1~mM H$_2$SO$_4$.
(a): 
CV profiles, d$E$/d$t = 2$mV/s. 
Experimental (dot-dashed) and simulated 
(30$\times$30: solid; 45$\times$45: $\times$) current densities. 
(b):
Simulated coverages of Cu (dashed), first-layer sulfate
(dotted), and total sulfate (solid), together with
corresponding Auger Electron Spectroscopy data ($\Box$ and $\times$ 
with error bars, respectively). 
After Ref.~\protect\cite{JZHA95B}. 
}
\label{figCVb}
\end{figure}

\begin{figure}[]
\caption[]{
Lattice-gas model for the UPD of copper (C) on Au(111)
in the presence of sulfate (S)
\protect\cite{JZHA95B}.
The relative positions of copper
($\bullet$) and sulfate ($\bigtriangleup$)
correspond to the effective interactions in Eq.~(\ref{eq1}).
The numbers are the corresponding values
of $\Phi_{\rm XY}^{(l)}$, in kJ/mol.
{}From Ref.~\protect\cite{RIKV96}. 
}
\label{figLATb}
\end{figure}

\begin{figure}[]
\caption[]{
Ground-state diagram for the lattice-gas model of copper
UPD on Au(111), shown in the
($\bar{\mu}_{\rm S} , \bar{\mu}_{\rm C}$) plane.
The effective interactions are given in Fig.~\protect\ref{figLATb}.
The solid lines represent zero-temperature phase boundaries.
The dotted line labeled ``1'' represents the voltammetric scan path at
room temperature, corresponding to the data shown 
in Fig.~\protect\ref{figCVb}. 
The lines labeled ``2'' and ``3'' correspond to 5~mM and 0.2~mM 
Cu$^{2+}$ with unchanged sulfate concentration, respectively.  
The end points of all three lines correspond
to $E$=120~mV (upper left) and 420~mV (lower right) vs.\ Ag/AgCl. 
The solid diamonds indicate the positions of Peak \#~1, and 
the solid squares indicate the positions of Peak \#~2 
in the simulated room-temperature CV currents. The phases are denoted as
$(X \! \times \! Y)_{\Theta_{\rm C}}^{\Theta_{\rm S}}$.
{}From Ref.~\protect\cite{JZHA95B}.
}
\label{figGSb}
\end{figure}

\begin{figure}[]
\caption[]{
Ground-state configurations corresponding to the main
phases in the ground-state diagram,
Fig.~\protect\ref{figGSb}. The adsorption sites are
shown as {\large $\circ$}, and Cu and sulfate are denoted
by {\large $\bullet$} and $\triangle$, respectively,
as in Fig.~\protect\ref{figLATb}.
Adapted from Ref.~\protect\cite{JZHA95B}.
}
\label{figUPDconf}
\end{figure}

\begin{figure}[]
\caption[]{
Simulated current transients for potential steps $\Delta E$ 
of different magnitudes and directions across the phase transition at 
Peak \#~1. The potential is changed along the scan path marked 
``3'' in Fig.~\protect\ref{figGSb}, and it takes the system between the 
ordered $(\sqrt{3} \times \sqrt{3})^{1/3}_{2/3}$ phase and the disordered 
low-coverage phase. 
The time $t$ is given in MC steps per spin (MCSS), and the current 
density in elementary charges per Au(111) unit cell and MCSS. 
(a): Positive-going potential steps, $\Delta E =$ 24, 28, 31, and 33~mV. 
(b): Negative-going potential step, $\Delta E = -15$~mV. 
}
\label{figQuenchCurrent}
\end{figure}

\begin{figure}[]
\caption[]{
Simulated adsorbate coverages versus time,
following positive-going potential steps. 
The data correspond to the current transients shown in 
Fig.~\protect\ref{figQuenchCurrent}(a). 
(a): Copper. 
(b): Sulfate.
}
\label{figPQuenchCoverage}
\end{figure}

\begin{figure}[]
\caption[]{
``Snapshots'' of configurations produced during a simulation of a 
$21 \times 21$ system following a positive-going potential step 
with $\Delta E = 24$~mV. 
Nucleation and growth of domains of the equilibrium low-coverage phase are 
apparent. The times shown are $t$=0, 100, 300, and 500~MCSS. 
}
\label{figSnapPlus}
\end{figure}

\begin{figure}[]
\caption[]{
Simulated adsorbate coverages versus time,
following a negative-going potential step with $\Delta E = -15$~mV. 
The data correspond to the current transient shown in 
Fig.~\protect\ref{figQuenchCurrent}(b). 
Copper (solid); sulfate (dashed).
}
\label{figNQuenchCoverage}
\end{figure}

\begin{figure}[]
\caption[]{
``Snapshots'' of configurations produced during a simulation of a 
$21 \times 21$ system 
following a negative-going potential step with $\Delta E = -15$~mV. 
Many locally ordered domains, separated by domain walls, are quickly formed. 
The subsequent equilibration requires the elimination of domain walls 
by lateral diffusion and is consequently very slow. 
The times shown are $t$=0, 10, 200, and 1000~MCSS. 
}
\label{figSnapMinus}
\end{figure}

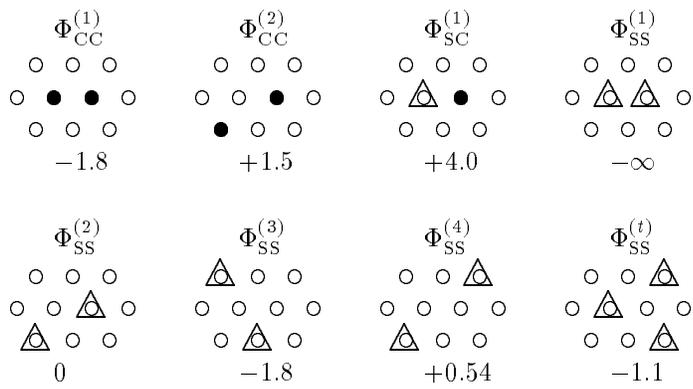
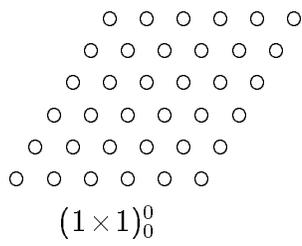
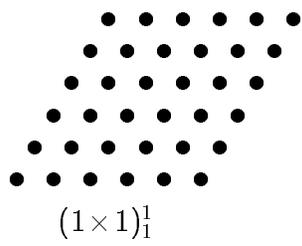
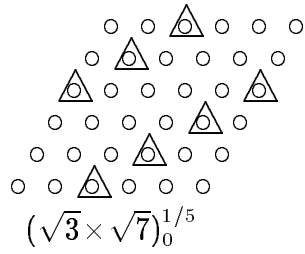
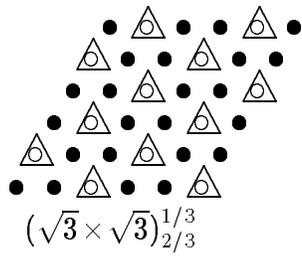
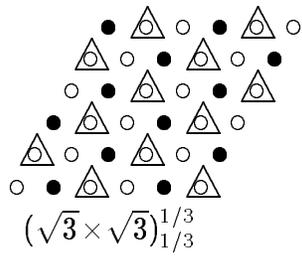
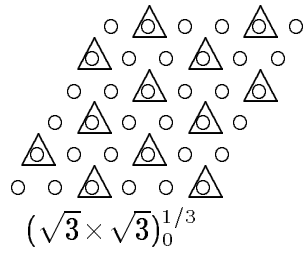
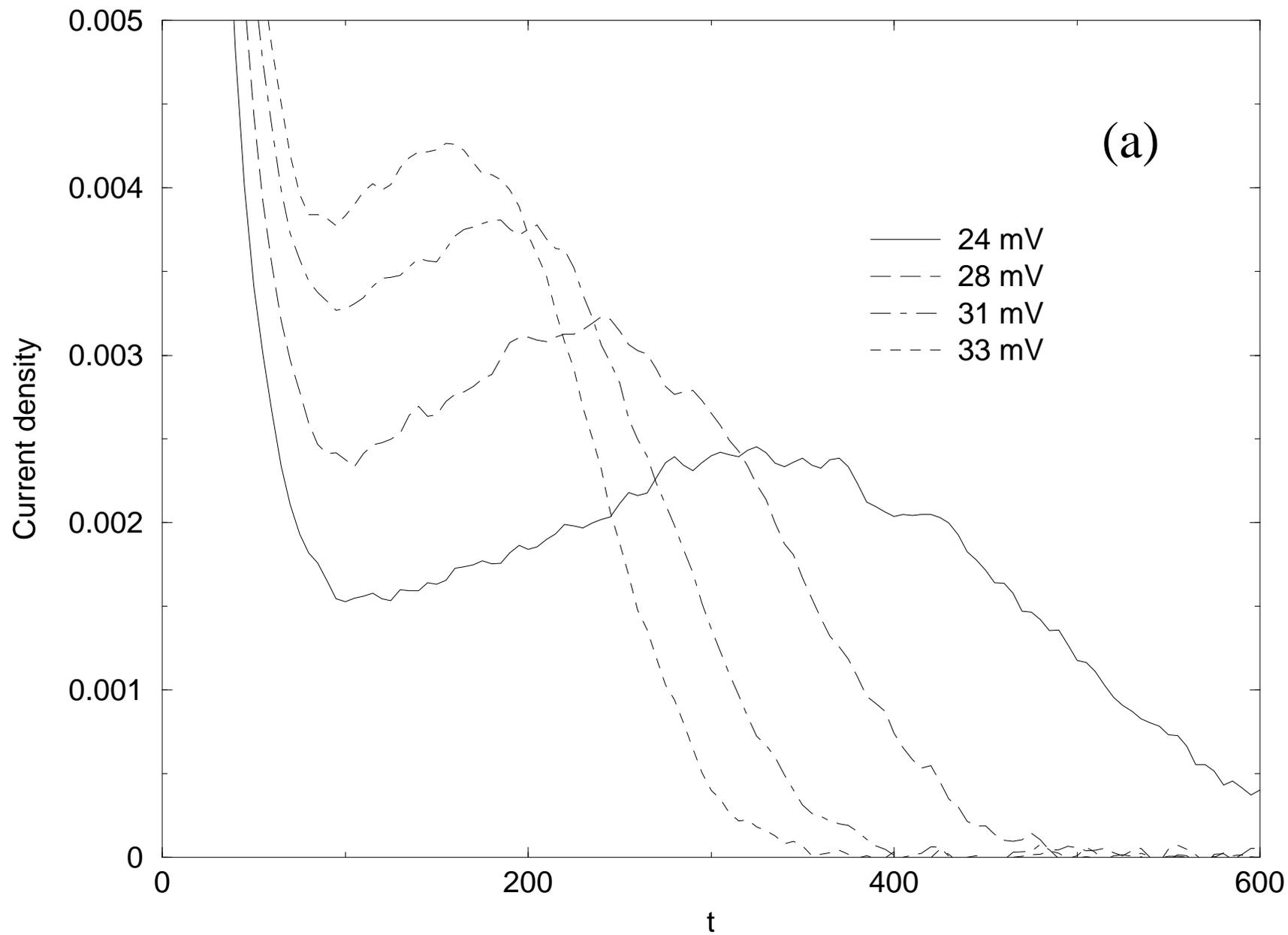
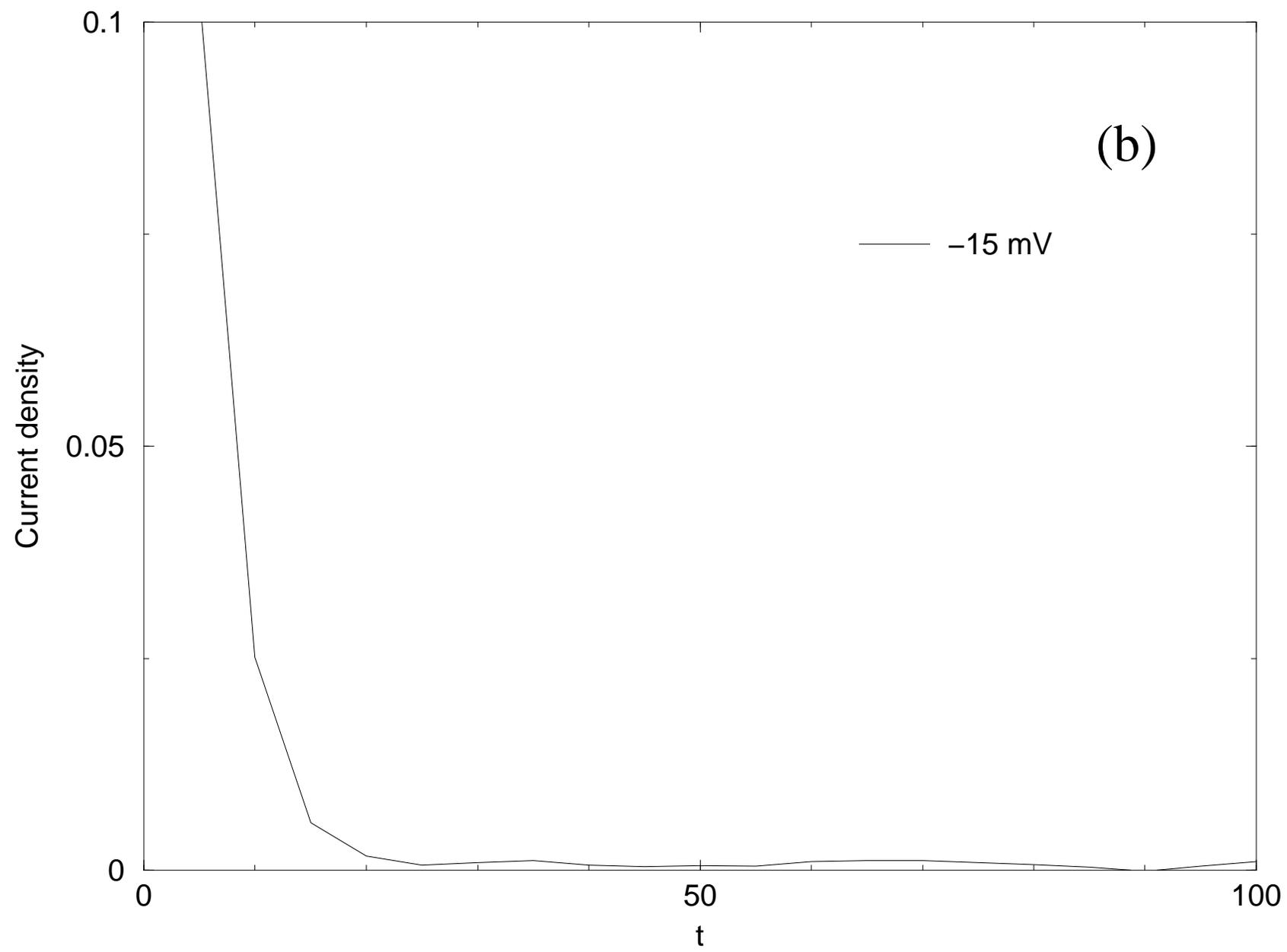
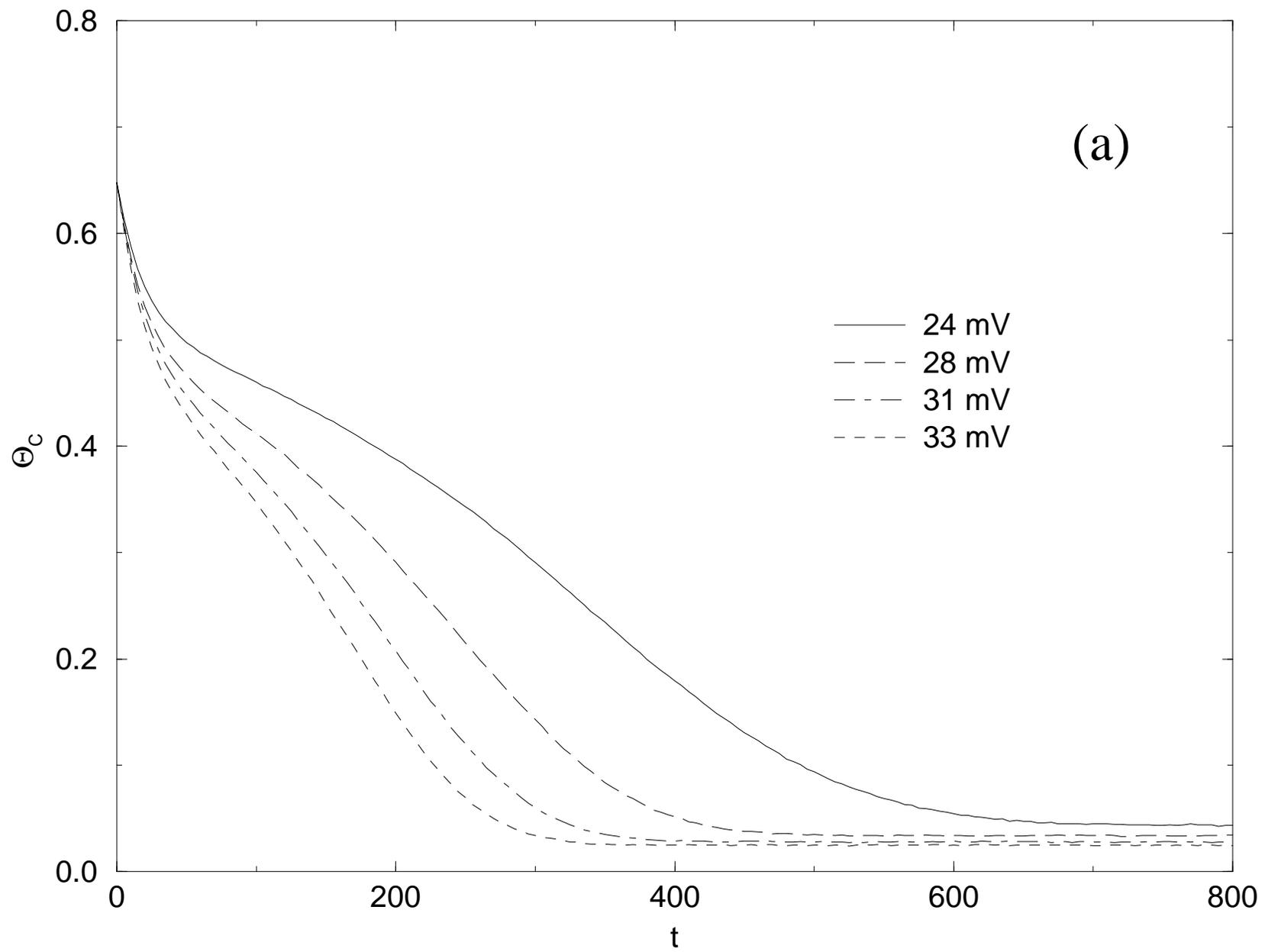
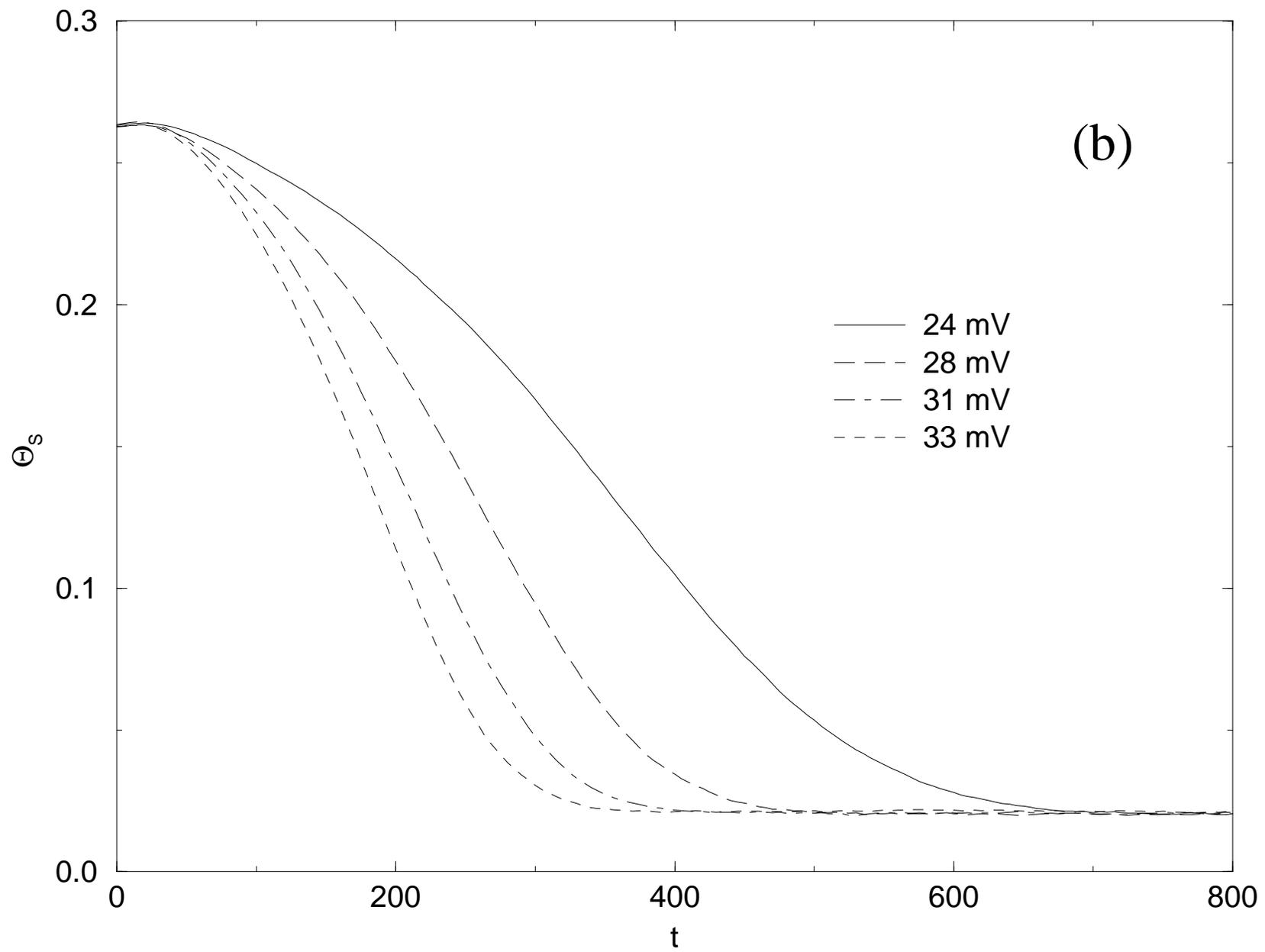
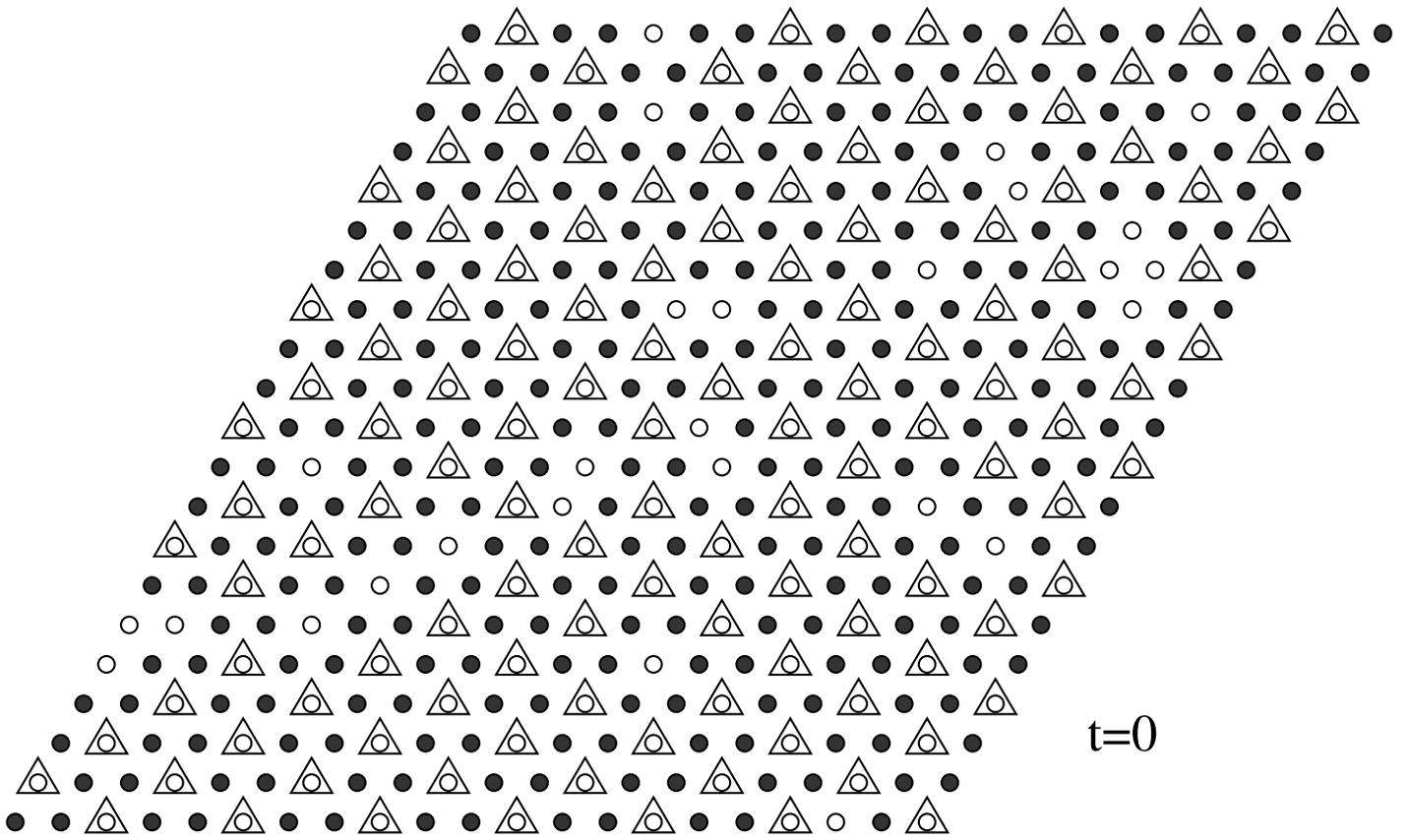
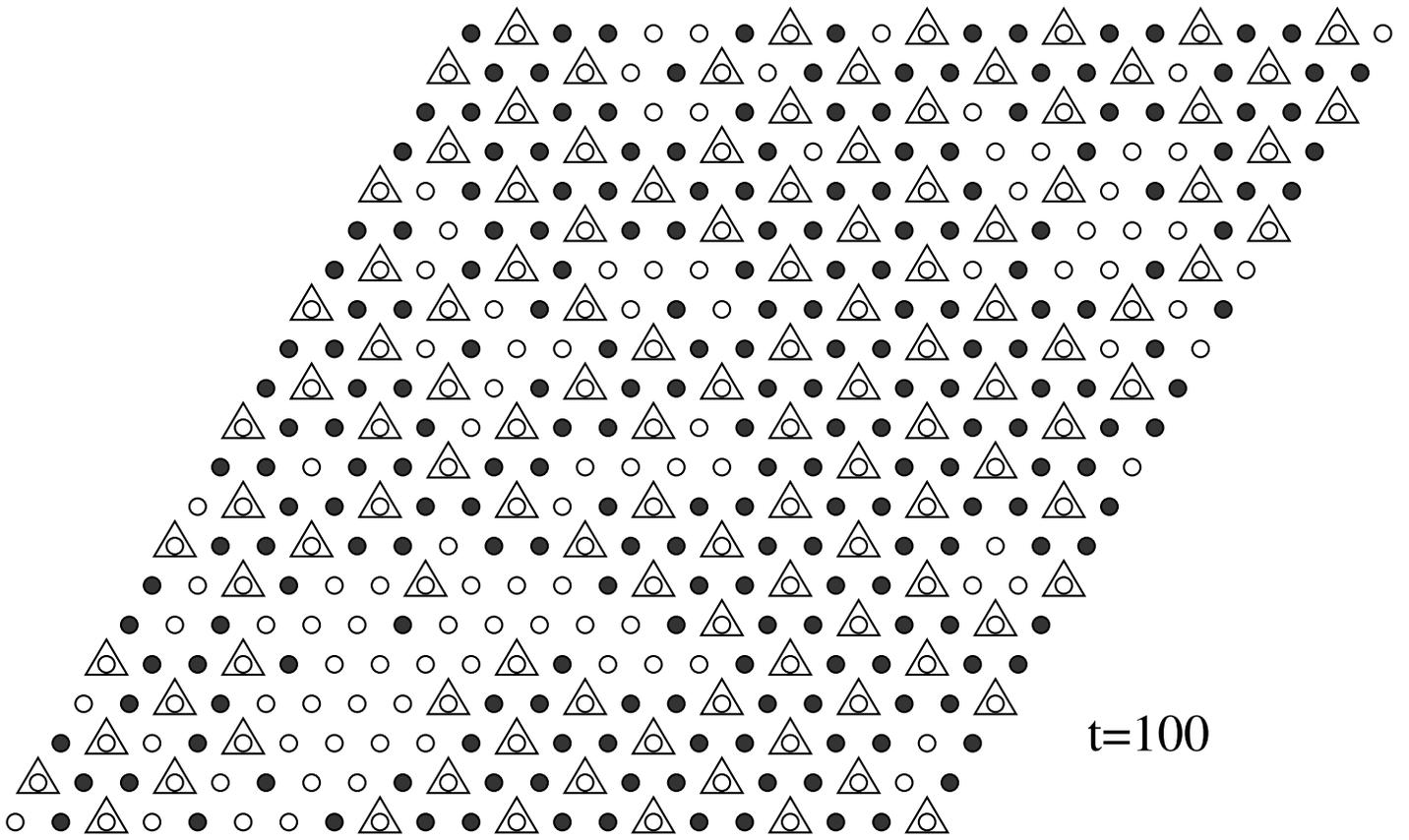
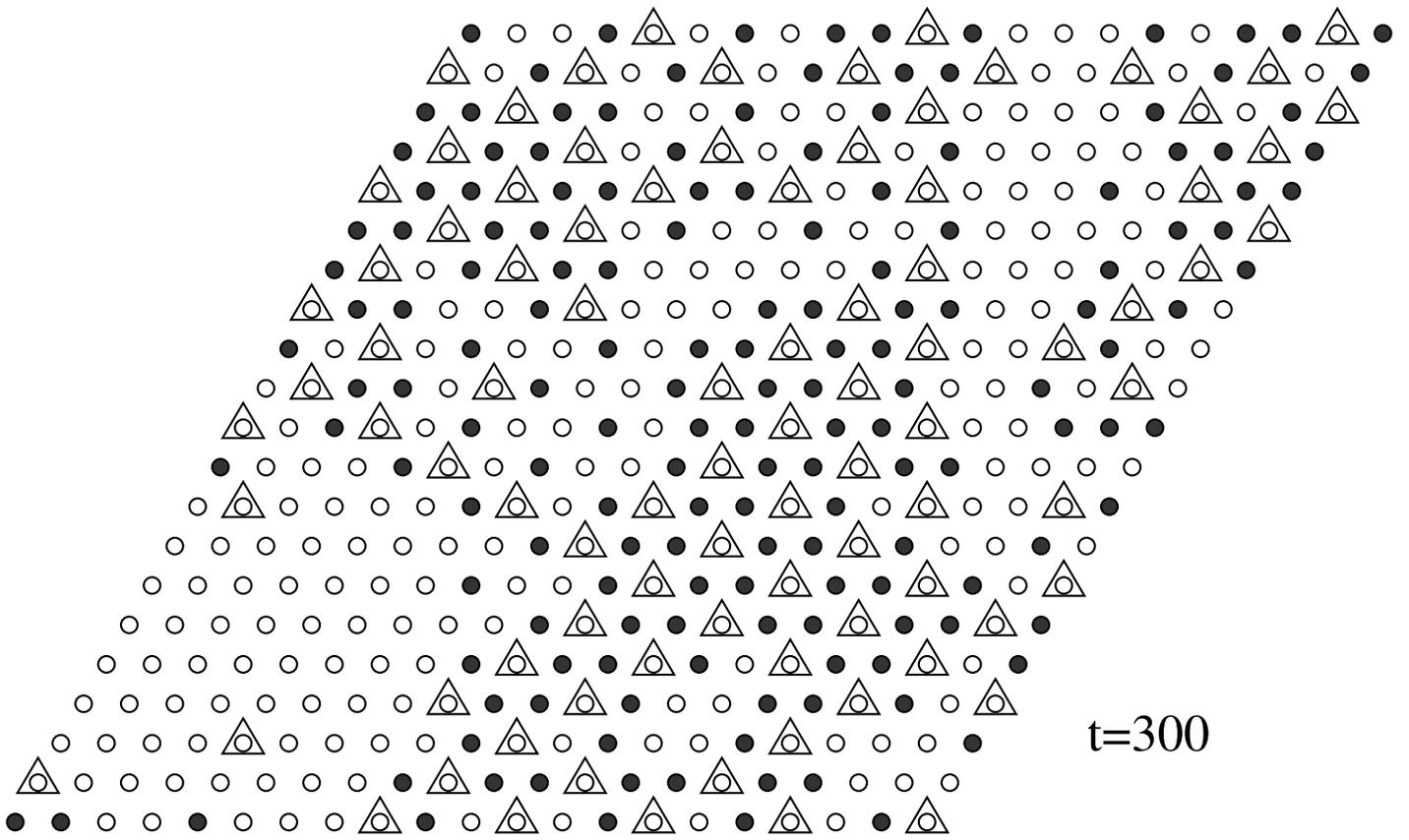
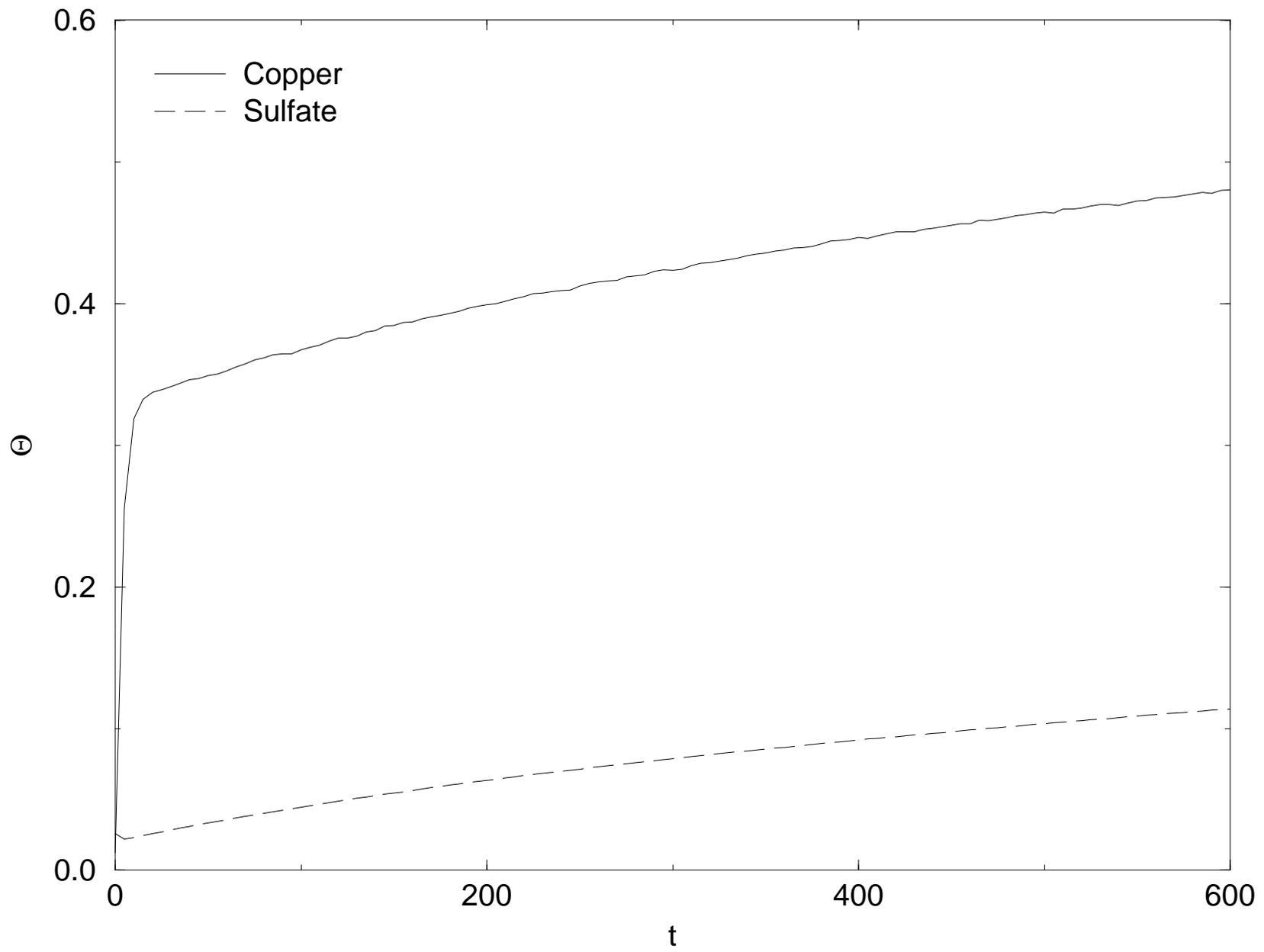
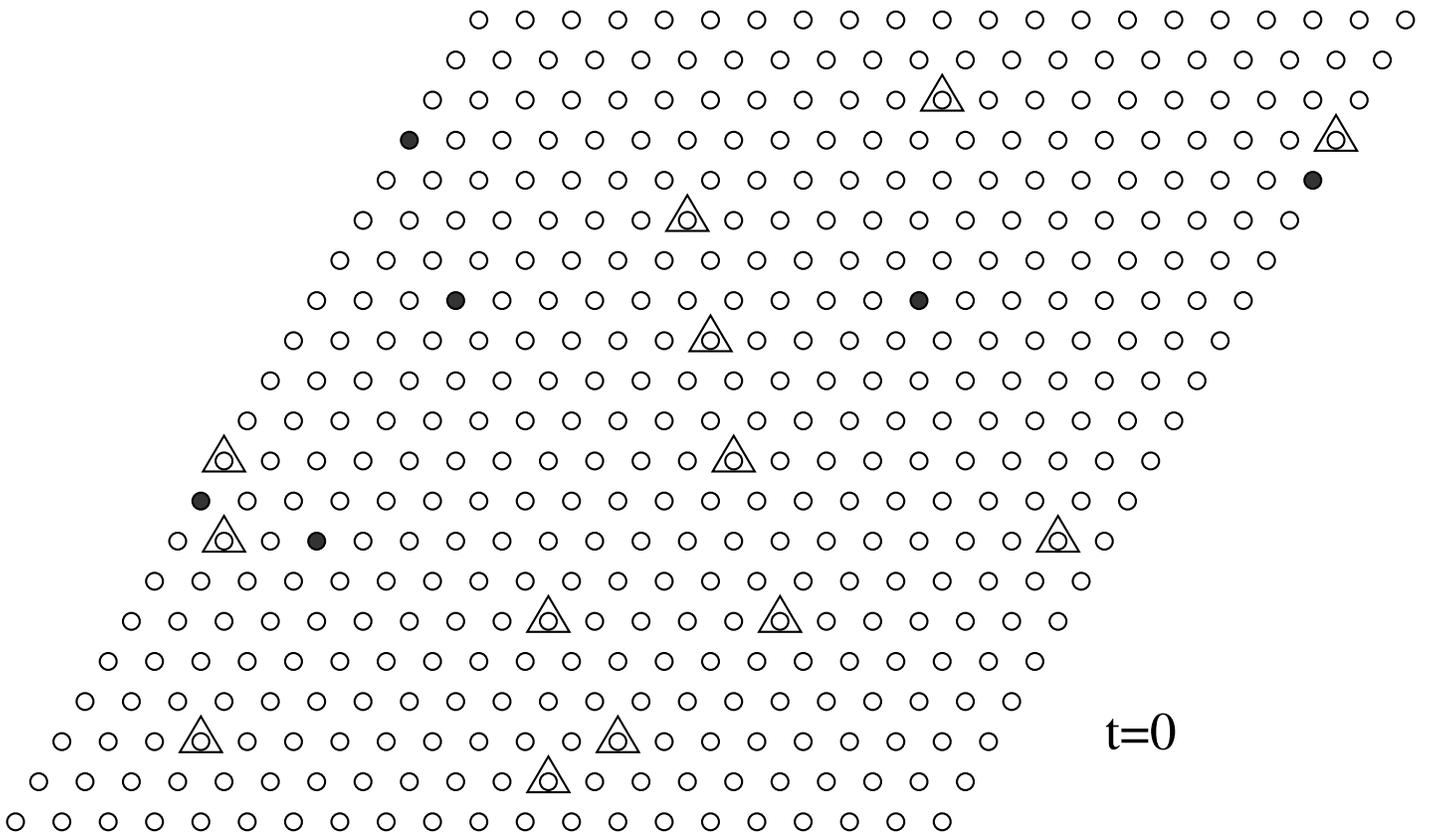
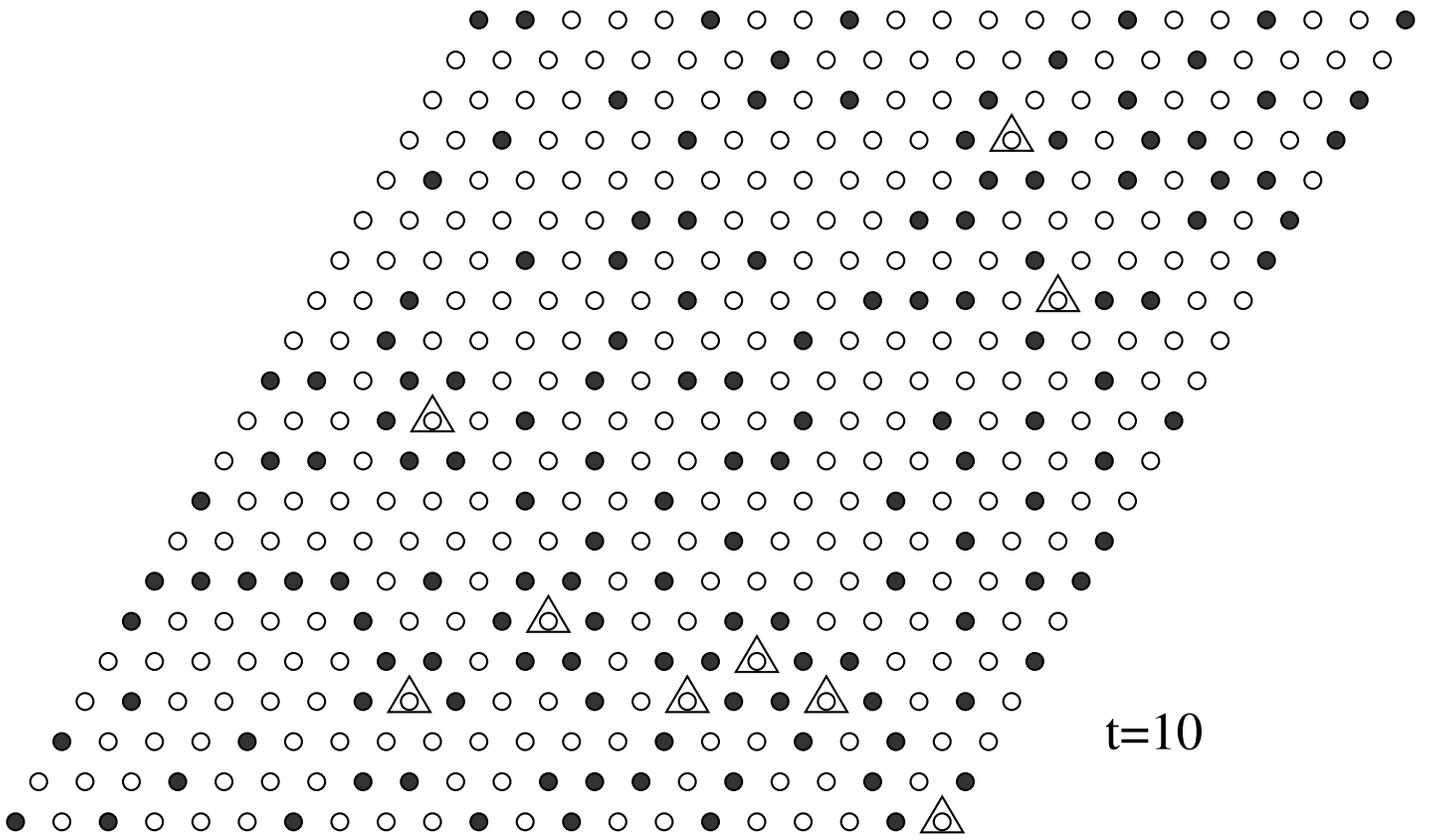
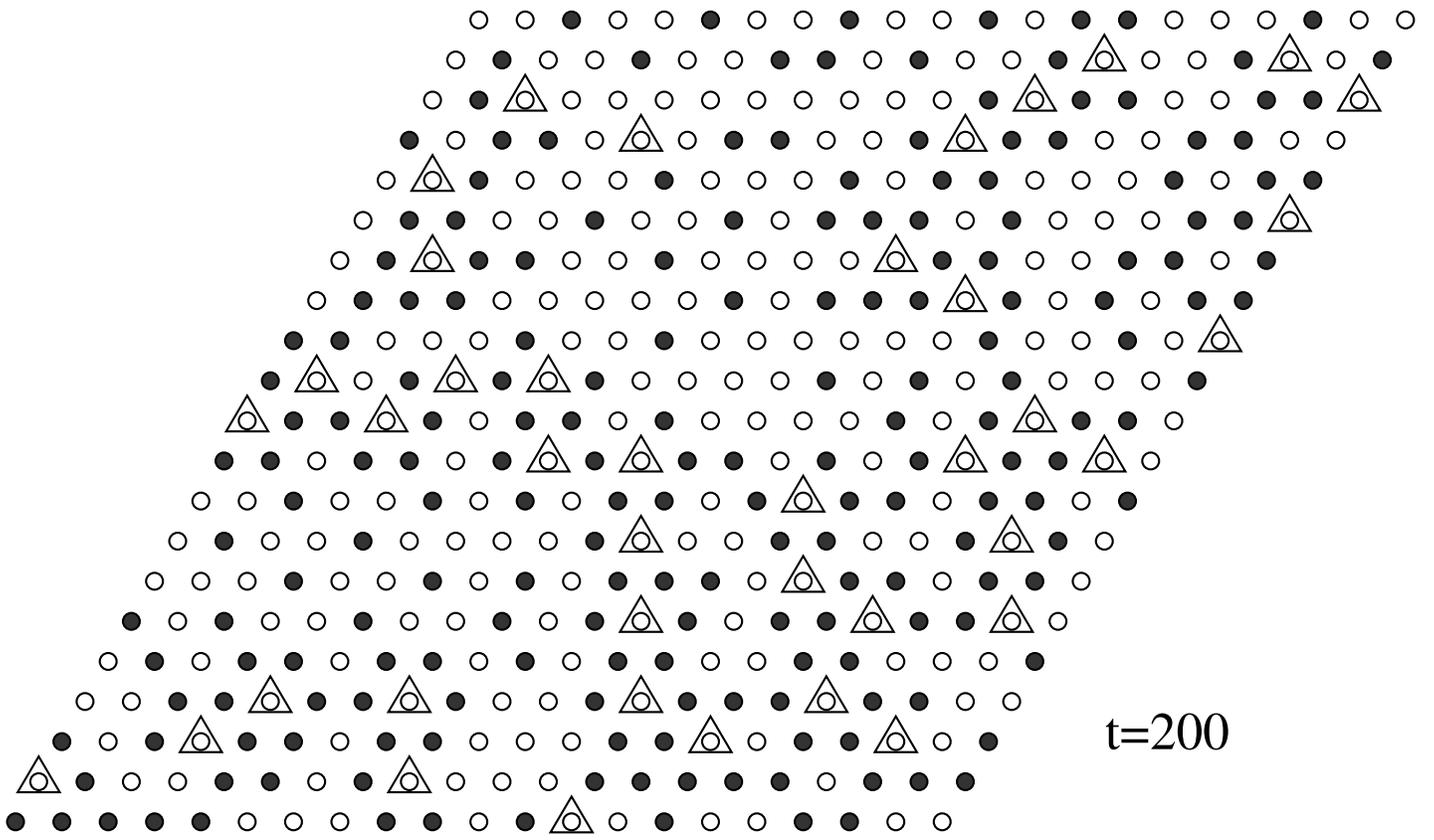
\end{document}